# Formation of high-aspect-ratio nanocavity in LiF crystal using a femtosecond of x-ray FEL pulse

Sergey S. Makarov,*[1] Sergey A. Grigoryev,[1] Vasily V. Zhakhovsky,[1] Petr Chuprov,[2] Tatiana A. Pikuz,[3] Nail A. Inogamov,[1,4] Victor V. Khokhlov,[1,4] Yuri V. Petrov,[4] Eugene Perov,[1] Vadim Shepelev,[2] Takehisa Shobu,[5] Aki Tominaga,[5] Ludovic Rapp,[6] Andrei V. Rode,[6] Saulius Juodkazis[7,8] Mikako Makita,[9] Motoaki Nakatsutsumi,[9] Thomas R. Preston,[9] Karen Appel,[9] Zuzana Konopkova,[9] Valerio Cerantola,[9,10] Erik Brambrink,[9] Jan-Patrick Schwinkendorf,[9] István Mohacsi,[9] Vojtech Vozda,[11] Vera Hajkova,[11] Tomas Burian,[11] Jaromir Chalupsky[11], Libor Juha,[11] Norimasa Ozaki[12] Ryosuke Kodama[12,13] Ulf Zastrau,[9] and Sergey A. Pikuz[14]

[1]Joint Institute for High Temperatures of Russian Academy of Sciences, 13/2 Izhorskaya st., 125412 Moscow, Russia.

[2]Institute for Computer Aided Design, Russian Academy of Sciences, Moscow, 123056 Russia.

[3]Institute for Open and Transdisciplinary Research Initiatives, Osaka University, Suita, 565-0871, Osaka, Japan.

[4]Landau Institute for Theoretical Physics of Russian Academy of Sciences, 1-A Akademika Semenova av., Chernogolovka, Moscow Region, 142432, Russia.

[5]The facility is Material Science Research Center, Japan Atomic Energy Agency, Sayo, Hyogo 679-5148, Japan.

[6]Laser Physics Centre, Department of Quantum Science and Technology, Research School of Physics, Australian National University, Canberra ACT 2600, Australia.

[7]Optical Sciences Centre and ARC Training Centre in Surface Engineering for Advanced Materials (SEAM), School of Science, Swinburne University of Technology, Hawthorn, VIC 3122, Australia

[8]Tokyo Tech World Research Hub Initiative (WRHI), School of Materials and Chemical Technology, Tokyo Institute of Technology, Tokyo 152-8550, Japan

[9]European XFEL, Holzkoppel 4, 22869 Hamburg, Germany.

[10]Università degli Studi di Milano Bicocca, Piazza della Scienza 4, 20126 Milano, Italy

[11]Department of Radiation and Chemical Physics, Institute of Physics, Czech Academy of Sciences, Na Slovance 1999/2, 182 00 Prague 8, Czech Republic.

[12]Graduate School of Engineering, Osaka University, Suita, 565-0871 Osaka, Japan

[13]Institute of Laser Engineering, Osaka University, Suita, 565-0871 Osaka, Japan
[14]HB11 Energy Holdings, Freshwater, NSW 2095, Australia.

**Abstract**

Sub-picosecond optical laser processing of metals is actively utilized for modification of a heated surface layer. But for deeper modification of different materials a laser in the hard x-ray range is required. Here, we demonstrate that a single 9-keV x-ray pulse from a free-electron laser can form a μm-diameter cylindrical cavity with length of ~1 mm in LiF surrounded by shock-transformed material. The plasma-generated shock wave with TPa-level pressure results in damage, melting and polymorphic transformations of any material, including transparent and non-transparent to conventional optical lasers. Moreover, cylindrical shocks can be utilized to obtain a considerable amount of exotic high-pressure polymorphs. Pressure wave propagation in LiF, radial material flow, formation of cracks and voids are analyzed via continuum and atomistic simulations revealing a sequence of processes leading to the final structure with the long cavity. Similar results can be produced with semiconductors and ceramics, which opens a new pathway for development of laser material processing with hard x-ray pulses.

**Introduction**

One of the areas of laser technology is the processing of materials based on substance ablation with ultrashort laser radiation in the optical, ultraviolet, and soft x-ray spectral range. The use of femtosecond pulses has several advantages over sub-nanosecond, nanosecond, and microsecond pulses, where material removal is mainly achieved by vaporization of the substance. In the case of femtosecond irradiation, the material is removed mainly by thermomechanical stresses[1–3], which makes it possible to significantly reduce lateral energy losses during laser drilling and improve the quality of the holes and cuts formed in the material. In optical laser

"drilling" technologies[4,5], the beam reaches the bottom of the cylinder to be formed and ablates the material at the bottom. In this case, the depth of the crater increases and becomes larger than the diameter of the beam on the surface. However, large length-to-diameter ratios cannot be achieved in this way (as a rule, this ratio does not exceed values in the order of one or few tens).

One possible way to increase this ratio is to generate microexplosion conditions with femtosecond pulses converted into a needle-like non-diffracting Bessel-Gauss beam. In essence, the Bessel-Gauss beam is cylindrically symmetric interference field created by the coherent superposition of cone-shaped optical plane waves[6]. It was shown that the Bessel-shaped pulse transforms much larger amount of a material by creating nanochannels with very high aspect ratio ~1:100[7,8]. This makes fs-Bessel pulses promising for creation and characterisation of warm dense matter (WDM) conditions leading to formation of new material structures. However, as in the case of microexplosion with Gaussian beams, formation of the Gauss-Bessel beams in the bulk of solid is possible only in transparent for the laser wavelength materials to attune the required $MJ/cm^3$ level of energy concentration.

The advent of free electron laser facilities (FEL) opens up unique opportunities for studying the effects of ultra-intense femtosecond XUV/X-ray radiation on various materials to determine the threshold of radiation resistance, model their decomposition, and secure further development technologies. Thus, in the work[9] it was shown that with the intense femtosecond irradiation of semiconductors (Si and GaAs) with an XUV FEL (FLASH - Free Electron LASer in Hamburg, $\lambda \sim 32$ nm) it is possible to form a crater with inner surface and boundary much smoother than those ablated by a conventional optical laser ($\lambda = 620$ nm). XUV/x-ray FELs represent promising tools for direct nano-patterning of solids, as they will enable the ablation, desorption, or phase transition printing of features with dimensions comparable to their short wavelengths. A key advantage of these lasers for direct fabrication of nano-structures is the unique combination of exceptionally short wavelength, high photon energy, high spatial coherence, and high peak power[10]. LIPSS (Laser-Induced Periodic Surface Structures) with a spatial period of 70 nm have already been produced on amorphous carbon[11] and poly(methyl methacrylate)[12] surfaces exposed to TTF1- FEL (TESLA Test Facility Free-Electron Laser, Phase 1) radiation at 98 nm and 86 nm, respectively. In a recent work [13], a new type of processing (the authors call it peeling) was developed when irradiating with femtosecond soft x-ray FEL pulses, where a threshold value is reached and the oxide crystal structure is destroyed. With a decrease in fluence F ~ 170 mJ/cm$^2$ and a photon excitation of 120 eV, the authors reported "exfoliation" of a substance leading to a formation of periodic nanostructures.

A limitation of the nano-structuring performed with XUV and soft x-ray FEL radiation lies in its relatively short attenuation length in numerous materials. New perspectives arise from the use of FELs that operate in the wavelength range of hard X-rays – XFELs (with wavelengths shorter than 1 nm), which have not previously been used in laser materials processing. It can be assumed that longer attenuation lengths of X-rays make possible to increase an aspect ratio of structures eroded by XFEL beams in a chosen material. In this context, the issue of studying damage and its thresholds for different materials when exposed to XFEL pulses with photon energies of several keV becomes relevant. These studies become possible thanks to the development of technologies for focusing XFEL beams to a spot of several tens to hundreds of nm (with a pulse energy of several tens to hundreds of µJ), which makes it possible to achieve an impact intensity on the sample in the range of ~ $10^{18}$—$10^{22}$ W/cm$^2$ [14–18].

The morphology of damage to a solid material exposed to X-rays depends on the intensity of the impact, the pulse duration and the energy of the incident photons. Conventionally, three main types of damage can be distinguished: (a) brittle damage in the form of cracks and chips, (b) ablative entry of the sample material in the vapor-plasma state into a vacuum in the spot and in the vicinity of the beam focusing spot on the surface of the irradiated sample. In our work, a new type (c) of damage to a substance (the so-called "radial extrusion" perpendicular to the beam propagation direction) under laser irradiation was obtained and described using a combination of experiments and computational and theoretical methods. The fact is that the transported substance deep in the material thickness cannot be transported over long distances along the channel axis to escape into the air or vacuum (the distances along the axis are large compared to the local value of the beam diameter). This situation differs fundamentally from the usual picture of laser ablation with the formation of a shallow crater whose depth is small compared to the beam diameter, see e.g.[1,19].

Here, we present the use of hard XFEL pulse ultrahigh intensity to create a deep cylindrical cavity with submicron diameter and aspect ratio no less than 1:1000 in the bulk of LiF crystal. LiF is often used as a fluorescent X-ray detector, which is ensured by the possibility of creating color centers (COs) when exposed to electromagnetic radiation with a photon energy of more than 14 eV[20–22]. In addition, this crystal is of interest due to its low atomic mass and relatively large mass difference between the constituting lithium and fluorine atoms (6.94 amu and 18.99 amu correspondingly). The large spatial separation of Li- and F-atoms due to the difference in the ion diffusion velocities is favorable for the formation of unusual crystalline structures of lithium, similar to bcc-Al formation observed in microexplosion experiments in sapphire [23]. The results of our simulations show that similar nanochannels with an extremely large length-to-diameter ratio can also be formed in metals, semiconductors, and ceramics, which is beyond the richness of optical lasers. It opens up completely unique possibilities for deep, highly targeted laser processing of materials.

**Experimental method**

The experiment was conducted at the High Energy Density (HED) instrument of the European X-ray Free Electron Laser (EuXFEL) [24]. An X-ray beam with a photon energy of 9 keV ($\lambda$ = 0.138 nm) and a duration of ~20 fs (full width at half maximum - FWHM) was focused through beryllium compound refractive lenses (CRLs) into a spot with a size of $d_{FWHM}$ = 0.41 µm, see Fig. 1A. A more detailed description of the focal spot size measurements for this experiment is given in[14]. The target was a circle LiF crystal with a diameter of 20 mm and a thickness of 2 mm and was put at the point of best focusing of the beam. Two regimes of single-pulse irradiation of a LiF crystal were investigated: $E_1$ = 26.9 µJ/pulse and $E_2$ = 81 µJ/pulse (the energy of the pulse was modified with an attenuation filter). Under these conditions, the absorbed energy density was $\xi_1$ = 297 kJ/cm$^3$ and $\xi_2$ = 895 kJ/cm$^3$, respectively (see Methods for conversion details). It is worth noting that these impact values are 2 orders of magnitude higher than the threshold value for LiF crystal damage (~ 4 kJ/cm$^3$ per pulse) determined in our previous work [25]. After each irradiation, the crystal was moved in a direction perpendicular to the beam so that the next irradiation would fall on the fresh surface of the sample.

Analysis of the interaction area of the XFEL beam with the LiF sample was performed with different readout systems (Figure 1B, see Methods for details). A scanning electron microscope (SEM) was used to study the damage on the surface, Fig. 1B (Step I). A confocal laser scanning microscope (LSM) in fluorescence mode was used to visualize the XFEL beam trace throughout the depth of the LiF, Fig. 1B (Step II). For additional analysis of deep damage near the LiF surface, the well-known FIB-SEM (Focused Ion Beam - Scanning Electron Microscopy) technology [26,27] was used, etching layer by layer deep into the LiF sample (the first few tens of micrometers from the surface along the XFEL propagation axis) with simultaneous visualization of the fracture morphology by SEM (45/90° to the XFEL beam), Fig. 1B (Step III).

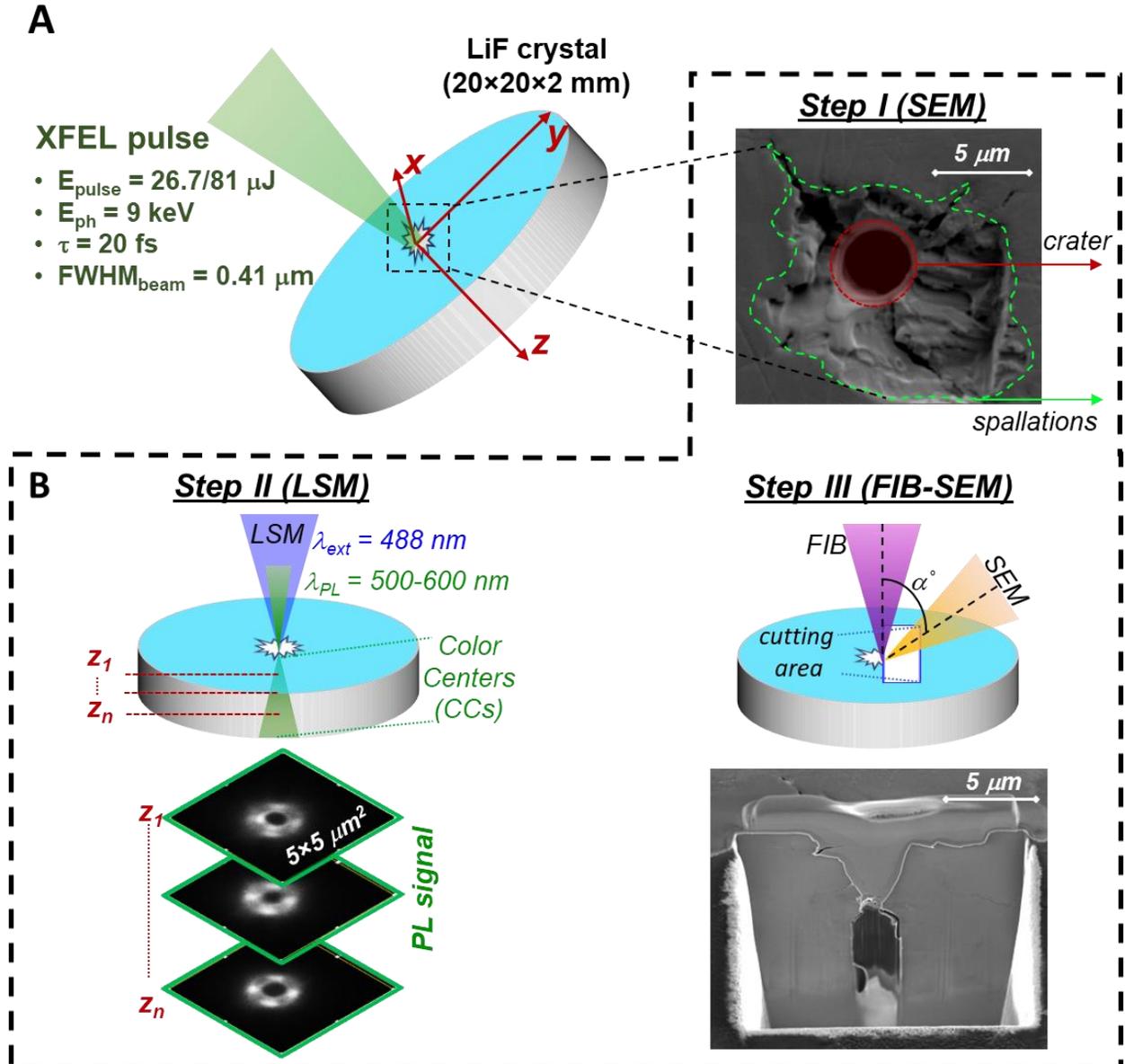

Fig.1 – **Experiment to create a cavity in a LiF crystal with an XFEL beam.** (**A**) - Scheme of the irradiation of a LiF sample. (**B**) – Diagnostics to study the morphology of the damage – step I (scanning electron microscope SEM), step II (reading the PL signal with a laser confocal microscope LSM), step III (study of the internal morphology of the cavity by cutting the sample with a focused ion beam and SEM illumination).

**Morphology of generated cavity**

*Surface morphology*

First, the surface of the exposed LiF sample in the area of interaction with the XFEL beam was analyzed with a scanning electron microscope (SEM). Figure 2(A-E) shows the scan results after a single-pulse X-ray exposure with $\xi_1$ = 297 kJ/cm$^3$ (A,B) and $\xi_2$ = 895 kJ/cm$^3$ (D,E). As can be seen from the images obtained by scanning with an electron beam at an angle $\alpha$ = 45° to the LiF surface (Fig. 2 A, D), the damage in both cases has a qualitatively similar structure: (1) a crater with clearly visible bulges of frozen material along its edges, (2) the area of delamination and spalling away from the influence of the X-ray beam. Viewing the damaged surface at an angle $\alpha$ = 90° (along the axis of incidence of the XFEL beam), Fig. 2 (B,E), reveals an additional feature in the damage pattern - a hole in the central area of the beam (wine-red area). The observed damage structure for two LiF irradiation regimes thus has the following qualitatively similar structure, Fig. 2(B,E):

1. Clearly visible hole in the area of maximum impact intensity $r_{hole-1}$ = 0.3 ×1 µm$^2$ ($r_{hole-2}$ =0.4×1.1 µm$^2$) - burgundy area. The indices 1 and 2 refer to $\xi_1$ and $\xi_2$.

2. Cylindrical melting region around a crater with a radius $r_{melt-1}$ = 0.7 µm ($r_{melt-2}$ = 1.65 µm) - purple region. We refer to the annular area between the wine-red and purple contours as a crater. We emphasize that both the melting region and the region of the hole entrance have a round shape. They are therefore caused by plastic processes and melting and not by a brittle fracture of the crystal.

3. Area of spalling of material with diverging cracks up to distances $r_{cracks-1}$ = 3 µm ($r_{cracks-2}$ = 10 µm) from the center of the beam.

If we compare the characteristic sizes of the damage for two irradiation energy densities, it becomes clear that the size of the entry into the cavity is approximately the same.

The distribution of the absorbed energy density $\xi$ of the incident beam on the surface of the LiF crystal is shown in Figure 2(C,F) for two irradiation regimes (see Methods for calculation details). Note that the distributions are plotted on a logarithmic scale up to the value $\xi = 10^{-2}$ J/cm$^3$ that corresponds to the threshold for redout the PL signal of $F_2$ and $F_3^+$ CCs with LSM [20,28]. As can be seen in Fig. 2(C,F), the decrease of $\xi(x)$ is quite strong - the level of intensity decrease $e$ times relative to the maximum of the Gaussian function is at a radius $r_{1/e} = FWHM/2/\sqrt{2} = 0.296$ μm. It is worth paying attention to the value of the absorbed energy density $\xi$ at the boundaries of the characteristic structures (2)-(3). To cause spalling of the material (case 3), the value of $\xi$ must be higher than the elastic limit LiF - 4 kJ/cm$^3$ [25] [blue horizontal line in Fig. 2(C,F)]. However, as shown in Fig. 2 (B-E), $\xi$ is much smaller than 0.01 kJ/cm$^3$ in the region of cracks. Consequently, the pressures directly associated with isochoric heating in these radii are much lower than 100 atm and these amplitudes are completely insufficient for the brittle fracture of a LiF crystal. A similar remark applies to the melting region (case 2) in Fig. 2 (B-E). Melting requires an increase in internal energy of about 6 kJ/cm$^3$ compared to the resting state at room temperature. At the edge of the melting range (purple contour), however, the local energy value is three orders of magnitude lower (~ 0.03 kJ/cm$^3$). Considering the above, the formation of regions (2)-(3) (melting region - crack region) is due to the attenuation of the shock waves (SW) generated by the influence of the XFEL beam on characteristic amplitudes.

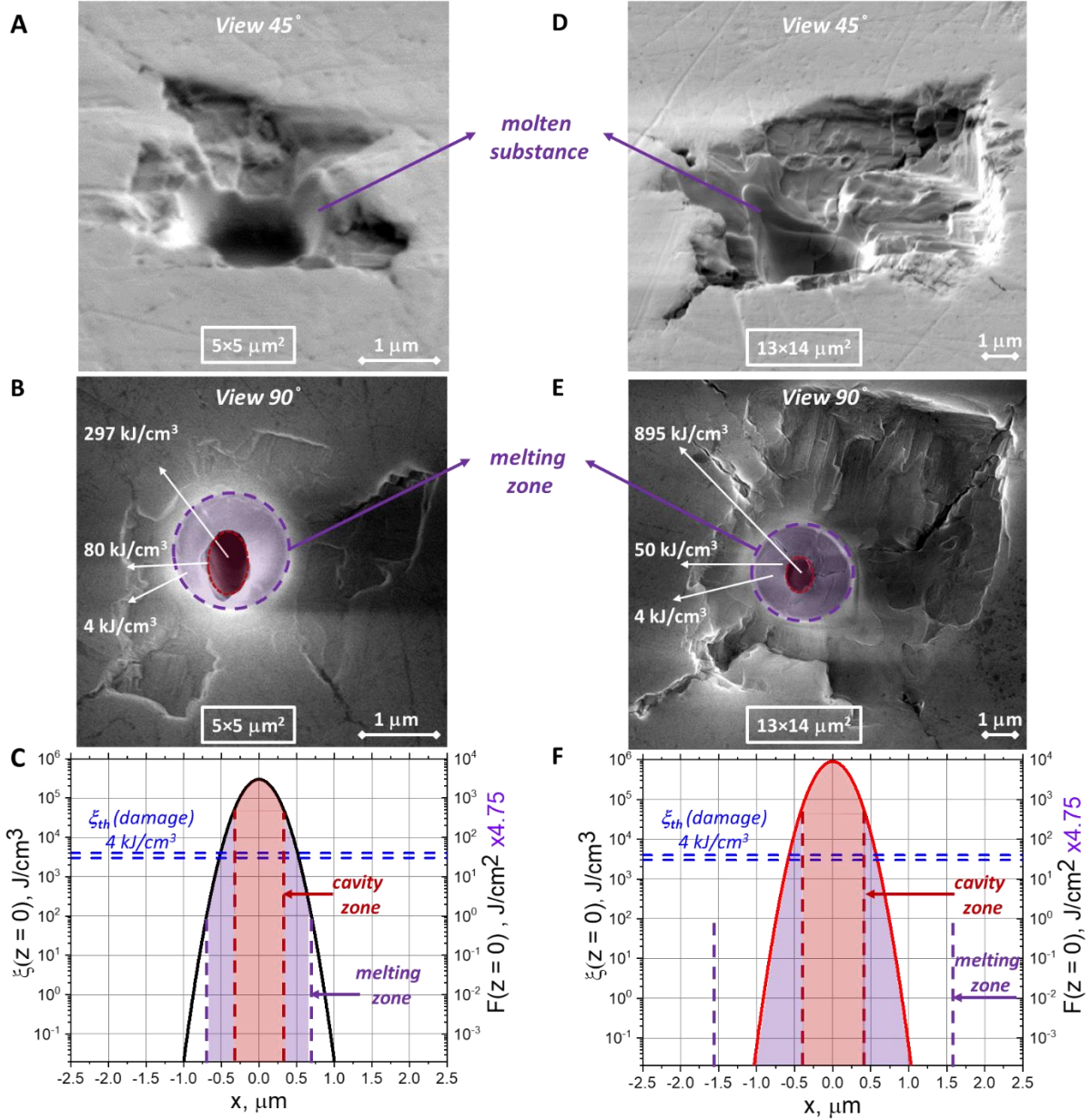

Fig.2 – **Morphology of the surface damage of an exposed LiF sample.** SEM image when viewed at an angle of 45-90° to the crystal surface after irradiation with an absorbed energy density of **(A,B)** — $\xi_1$ [26.9uJ/pulse]= 297 kJ/cm$^3$ and **(D,E)** — $\xi_2$ [81uJ/pulse] = 895 kJ/cm$^3$. **(C,F)** — Corresponding spatial distributions of the absorbed energy density on the surface of the LiF sample ($z = 0$). The purple and burgundy dashed lines correspond to the position of the corresponding regions highlighted in **(B,E)**.

*Damage at sample depth*

At the next stage, the structure of the internal damage (deep in the sample) was investigated, namely the hole observed at the bottom of the crater (wine-red area in Fig. 2(B,E)). To investigate the internal structure of the damage, we used a confocal laser scanning microscope (CLSM) with a layer-by-layer reading of the PL signal deep in the LiF sample (up to $z = 1200$ μm) with a step $\Delta z = 1.4$ μm (see details in Methods). Figure 3A shows the internal damage structure of a LiF crystal imaged with an LSM700 confocal microscope after a single pulse exposure with $\xi_1 = 297$ kJ/cm$^3$. PL images show the presence of a formed crater up to ~2 μm. Moreover, up to $z \sim 10$ μm, no crater is seen in the center area, but a PL signal is observed, indicating the presence of matter in this region. Note that since the first analysis of the sample surface with the SEM, we see a bright PL signal outside the X-ray beam region in the first layers of the LiF images (Fig. 3A - $z = 2$-8 μm), which is explained by the excitation of color centers by the electron beam in the first layers of the sample.

An effect in the form of the absence of the PL signal in the central region of the beam is observed for depths $z$ more than ~ 10 μm and up to $z \sim 1000$-1100 μm, after which the PL signal appears again, Fig. 3A. Thus, a long channel is formed deep in the LiF sample. In Figure 3B, the black dots show the measured channel widths as a function of the LSM scan depth in the LiF sample. We can see that the damage of LiF forms a complex structure after a single pulse irradiation with a beam width of $d_{FWHM} = 0.41$ μм и $\xi_1 = 297$ kJ/cm$^3$: crater (~ 2 μm); layer of matter ("plug" ~ 7 μm); an ultra-deep channel (length $L_c \sim 1000$-1100 μm) with a diameter $d_c \sim 0.6$ μm throughout the entire depth.

The formed channel in LiF has a very high aspect ratio ($d_c:L_c \sim 1:1600$-1800). A similar damage structure was observed for the irradiation regime with $\xi_2 = 895$ kJ/cm$^3$, Figure 3 (B-red dots), but with a slightly larger channel width and length. Due to the limitation of the working distance of the confocal microscope we could not observe the depth at which the channel disappears in this case.

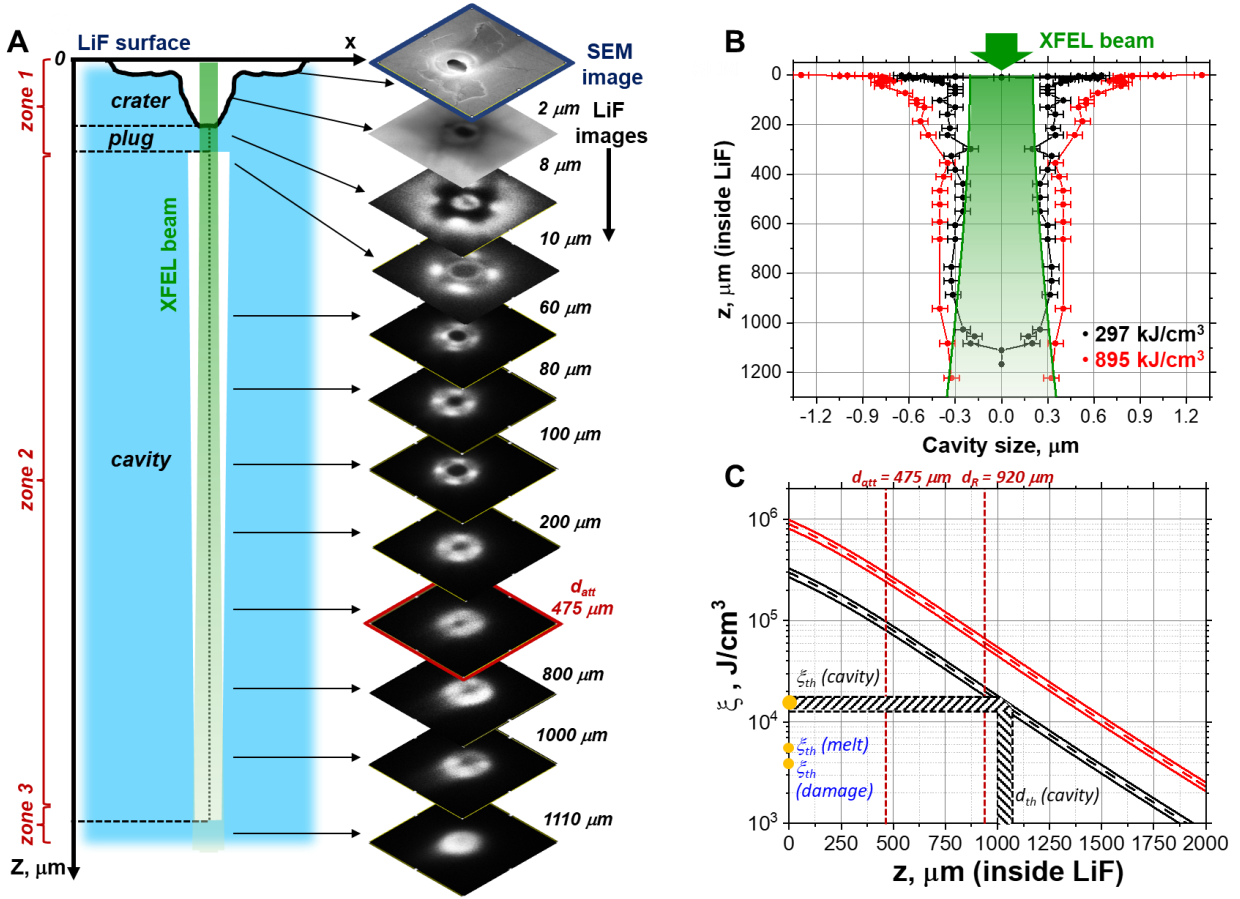

Fig. 3 – **Investigation of the internal structure of LiF destruction using a confocal microscope.** **(A)** ─ Readout data of an exposed LiF crystal at an absorbed energy density $\xi_1 = 297$ kJ/cm$^3$. The readout was performed with an LSM700 confocal fluorescence microscope deep inside the crystal in the range $z = 0$ μm (surface) – $z = 1200$ μm. The identified areas within the sample are shown schematically on the left. **(B)** ─ cavity dimensions measured from PL images as a function of LiF scan depth for exposures with $\xi_1 = 297$ kJ/cm$^3$ (black dots) and $\xi_2 = 895$ kJ/cm$^3$ (red dots). The green shading shows the region of propagation of the XFEL beam deep in LiF, taking into account the divergence (calculation details in Methods). **(C)** ─ Distribution of the absorbed energy density $\xi$ within the LiF crystal at different depths $z$ (in the direction of the XFEL beam) after a single pulse: the black line corresponds to $\xi_1 [z = 0] = 297$ kJ/cm$^3$ at the surface, and the red line corresponds to $\xi_2 [z = 0] = 895$ kJ/cm$^3$.

Figure 3C shows the dependence of the decay of the absorbed energy density $\xi$ within the LiF crystal for two irradiation modes ($\xi_1 = 297$ kJ/cm$^3$ ─ black curve, $\xi_2 = 895$ kJ/cm$^3$ ─ red curve) (calculation details in Methods). The value of $\xi_1[z = 1000\text{-}1100\ \mu m] = 12\text{-}18$ kJ/cm$^3$, where the channel ends, is ~16-25 times lower than at the surface $\xi_1[z = 0\ \mu m] = 297$ kJ/cm$^3$. The found threshold for channel formation $\xi_{th}$(damage) ~3-4 times higher than the threshold for the onset of LiF dielectric damage found in our previous work[25].

To verify that the observed absence of the PL signal in the LiF images in the central region of the XFEL beam in Fig. 3A is due to the absence of matter (cavity), we cut the LiF sample layer by layer near the expected cavity to a depth of a few tens of micrometers from the surface (see Methods for details). We used FIB-SEM technology, in which an ion beam ablated the LiF sample layer by layer and a scanning electron microscope (SEM) visualized the fracture morphology (at an angle of 45° to the channel plane), see Fig. 1B (Step 3). Figure 4(A,B) shows the results of the corresponding SEM measurements after crystal etching for an exposure with $\xi_2$ = 895 kJ/cm³. In Fig. 4A, with a wide field of view, the existence of a cylindrical cavity under the crater bottom is observed. With further etching of the LiF material and a narrower scan range, the "crater-initial channel" boundary is clearly visible, Fig. 4B. The crater has the shape of a cone with a base size of ~ 3.3 μm, which is consistent with the melting area $r_{melt-2}$ = 3.3 μm in Figure 2(E,F). The images in Fig. 4(A,B) show a similar fracture structure as the PL images in Fig. 3A: (1) crater (cone-shaped), (2) thin substance layer (plug), (3) expanded cavity. According to the SEM images for the case of exposure with a LiF sample with $\xi_2$ = 895 kJ/cm³, Fig. 4B, the cavity width decreases from 2.8 to 2 μm at depths up to $z$ = 21 μm, which also corresponds to the measurements of this value from the PL images in Fig. 3(B – red dots). At greater depths, the channel narrows to a radius of 0.4 μm and its size remains practically unchanged (Fig. 3(B – red dots)).

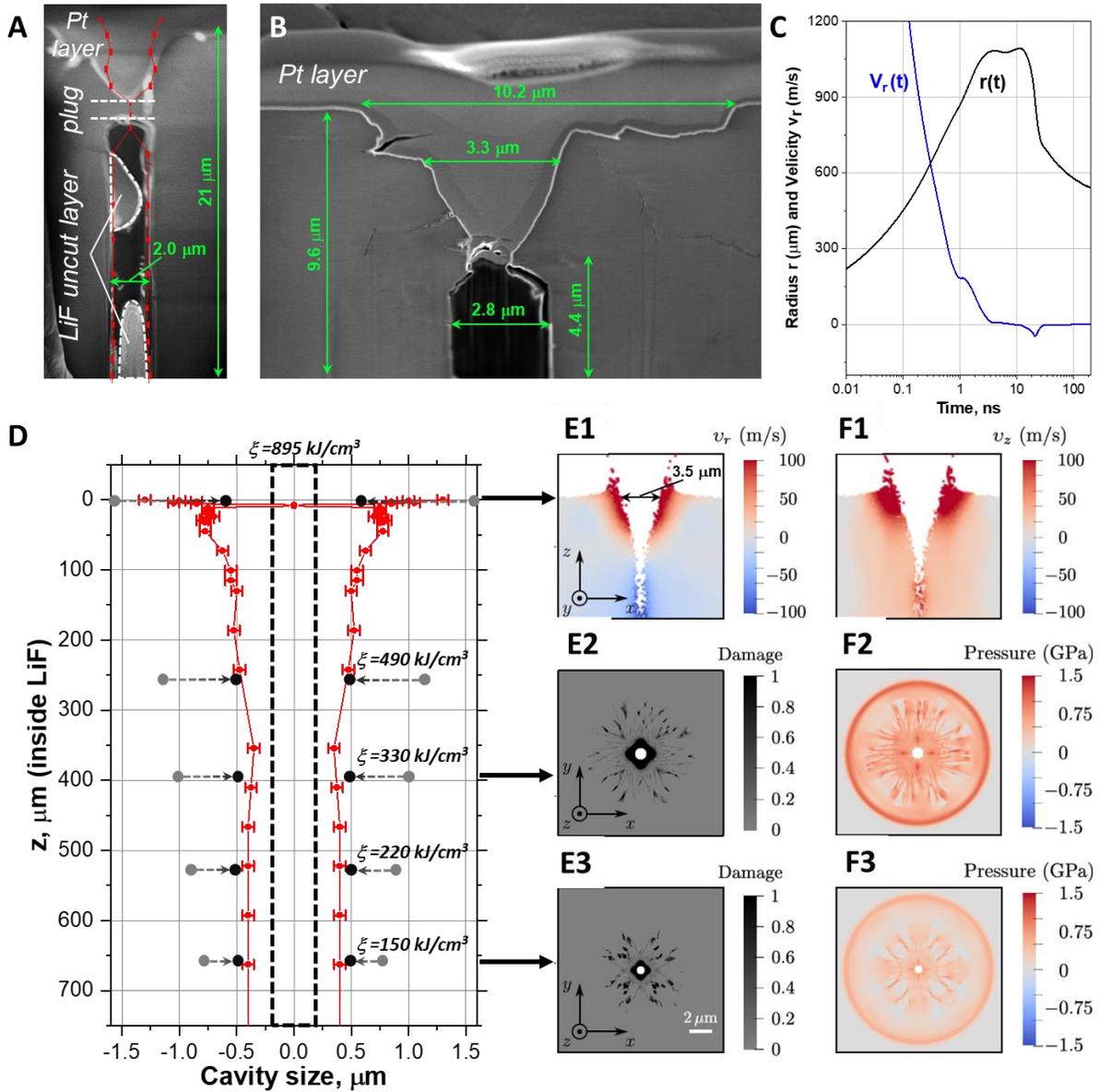

Fig. 4 – **Comparison of experiment and SPH simulation.** FIB-SEM measurements of a LiF crystal irradiated with $\xi_2$= 895 kJ/cm$^3$ in one pulse: **(A)** — wide scan area near the crystal surface (PL measurements from Fig. 3B are shown in red). **(B)** — narrow area at the "crater-graft-beginning of the cavity" boundary as the crystal is further cut. Results of modeling the formation of a cavity in the irradiation mode $\xi_2$ = 895 kJ/cm$^3$: **(C)** — Evolution of the cavity radius $r(t)$ and the radial velocity of its boundary $v_r(t)$ at a depth of 390 μm (2D). **(D)** — Dependence of the final cavity size on depth: red dots correspond to experimental data; black and gray dots correspond to simulation results (2D). The initial cavity size given in the simulations is indicated by two vertical dashed lines. The gray circles correspond to the maximum radius to which the cavity expands in the simulations. **(E1, F1)** — Simulated fields of radial $v_r$- and axial $v_z$-velocities formed in the $xz$-plane near the surface (3D). **(E2, E3)** — LiF damage patterns formed at a depth of 390 and 660 μm at time t ≃ 1.75 ns after the start of irradiation (2D). The corresponding pressure field is shown on the right in the insets **(F2, F3)**. The size of the images is given on a scale of 15 × 15 μm$^2$.

The experimental results on the formation of a deep cavity with an ultra-high aspect ratio inside the LiF material require a theoretical explanation, as such a thing has not been observed before. Since the formed structure has a complex morphology and a high aspect ratio (see Fig. 3A - zones 1-3), a quantitative description with only one model and one numerical code cannot describe the whole physics of the formation of such a structure. In this context, the modeling areas were divided into the following (see Fig. 3A): (zone 1) crater near the surface and beginning of cavity formation (3D SPH, 2D HD); (zone 2) substance plug and long cavity (2D SPH, MD); (zone 3) area of disappearance of the cavity in the depth of the LiF sample (2D MD).

**Multi-method simulations**

The aim of the numerical simulation is to relate the energy release of the XFEL beam to the features formed (crater, plug, cylindrical cavity radius, channel floor). In the following simulations, we used the distribution of the value of ξ inside the LiF crystal, as shown in Fig. 3C.

*Crater formation and axisymmetric expansion of LiF in the depth of crystal*

To simulate the dynamics of cavity opening, as well as damage to lithium fluoride as a result of irradiation, the LiF model implemented within the framework of the smoothed particle method[29,30] (see details in Methods) is used as in our previous work[25]. It provides a detailed description of the equation of state and failure model for LiF, and presents results obtained at moderate radiation intensity, near the material damage threshold.

Figure 4C shows the simulated curves of the cavity radius $r$ and the radial velocity of its boundary $v_r$ as a function of time. The diagram shows that the cavity expands and reaches its maximum size in the first few ns after the start of irradiation. At time $t \sim 3$ ns, the growth of the cavity practically stops, as can be seen from the velocity curve $v_r(t)$. After that, its size remains practically unchanged, as in the experiment.

The contraction of the cavity begins after the arrival of a rarefaction wave from the free boundary of the target, which reduces the pressure inside the cavity to zero. To account for the effect of this rarefaction wave in our 2D simulations, the pressure inside the cavity starts to drop after a certain time corresponding to the depth of the simulation layer. In our simulations, the pressure drop occurs within ~10 ns. After the pressure has dropped to zero, elastic forces around the cavity lead to a contraction of its radius. The diagram in Fig. 4C shows that in the time interval $t = 10-20$ ns the radial velocity $v_r(t)$ becomes negative and at the same time the radius of the cavity r(t) begins to decrease sharply. After the pressure inside the cavity has completely dissipated, at $t > 20$ ns, the cavity continues to narrow due to inertia, but at a much slower speed. The diagram of the radial velocity $v_r(t)$ shows that it approaches zero at $t > 30$ ns, i.e. the change in cavity size practically comes to a standstill. From this moment onwards, the radius of the cavity tends towards its asymptotic value.

Figure 4(D) shows a comparison of the cavity radius as a function of depth, which was determined in the experiment and in the simulation in irradiation mode $\xi_2 = 895$ kJ/cm$^3$. Firstly, there is a good quantitative agreement between the simulations and the experimental data. Secondly, you can see from the above diagram that in the simulations the final size of the cavity is approximately the same at different depths (it decreases slightly with depth), although the maximum achievable size varies greatly. The experiment also shows a similar picture: the transverse size of the cavity remains practically unchanged over the entire depth of the channel. A noticeable widening is only observed near the free surface.

In the inserts (E2,E3) to Figure 4 shows the image of LiF damage obtained in the simulations for the irradiation mode $\xi_2$ = 895 kJ/cm$^3$ at different distances from the irradiated surface of the sample — depth 390 μm (E2) and 660 μm (E3). The gray color in the figures represents the original, undamaged material, the black color corresponds to the damaged area. The central part, shown in white, is a cavity. From the pictures above you can see that the damaged area decreases with depth. This is quite natural, as the specific energy absorbed and therefore the amplitude of the shock wave decreases with depth. The area of continuous damage is followed by an area of partial damage. Here, the material is damaged along clearly defined damage bands, the concentration of which gradually decreases with the distance from the area of initial heating. At a distance of 5−10 μm from the irradiation point (depending on the depth), the damage to the material stops.

To analyze the dynamics of the cavity opening near the surface of the LiF sample, the modeling was performed in a 3-dimensional environment. In contrast to the expansion of the cavity in depth, here the type of deformation is mainly determined by the relief of the resulting stresses on the free surface of the sample. In the insets (E1, F1) to Figure 4, the radial $v_r$ and axial $v_z$ velocities are shown in longitudinal section. From the above figures, one can recognize the beginning of the formation of a cone-shaped crater similar to the one observed in our experiments (see Fig. 4B). From the velocity distribution $v_r$ shown in Fig. 4(E1), it is clearly visible that the crater continues to expand near the surface, while at depth the cavity has already started to contract (the blue color in the lower part of the figure (E1) corresponds to the negative radial velocity). The underestimation of the final radius determined in 2-dimensional simulations for the case of cavity expansion near the surface (Fig. 4(C)) therefore seems reasonable, as the influence of the free surface was not considered. In a 3-dimensional simulation, where such an influence should be considered naturally, the radius of the crater is many times larger than the radius of the cavity at depth.

It is also noteworthy that the damage area determined in the simulations has a cruciform shape similar to the experimental images of a confocal microscope, see Fig. 3A and Fig. 4(E2,F2). The reason why the damage area forms a structured cruciform shape in the simulations is that the SPH particles initially occupy the nodes of a close-packed 2-dimensional lattice. In this formulation, there are selected directions along which the damage to the LiF occurs. A similar picture that emerges in the experiment may also be since the sample contains selected directions related to the single crystallinity of the investigated samples.

*MD simulations*

To explain the mechanism of formation of the cross-shaped structure visible in the PL images (see Fig. 3A) around the cavity at the sample depth $z > 8$ μm, the large-scale MD simulation of the radial material movement in a deep slice of LiF crystal perpendicular to a rapidly heated cylindrical channel was performed (see Methods for more details). The high pressure in the central hot spot (channel cross-section) generates a diverging cylindrical shock and a converging rarefaction wave, which both causes a radial flow of material leading to decrease of plasma pressure and temperature within the channel. As in the SPH simulation, the surrounding cold crystal is damaged by a strong shock wave until it weakens enough. Due to the cubic crystal structure of the crystal, four corner damage sectors are formed around the radial cracks, which can be seen in Fig. 5A. After traveling a sufficient distance from the central axis, the diverging shock wave attenuates and becomes purely elastic. Then the irreversible damage ceases and the cold solid begins to elastically resist the radial outflow of material from the channel. Soon the flow velocity reduces to zero, and the remaining pressure of about 1-2 GPa in the hot fluid within the channel comes to the mechanical equilibrium with the tightening elastic stress in the surrounding cold solid.

Thus, the cylindrical cavity is still not formed in the channel filled by low-dense hot fluid. At this point, the rapid (acoustic) stage of material transformation ends and a long cooling stage of the channel fluid leading to liquid-vapor phase separation and crystallization of the melt begins.

Figure 5A shows the 2D distributions of the damage and the equivalent von Mises stress after the damage has ceased, as well as the experimental PL image at a depth of $z = 10$ μm. It is important to underline that the bright, cross-shaped PL signal cannot be caused by the XFEL beam being angularly uniform about its axis. We believe that such an angular dependence of the concentration of color centers in LiF could arise under the influence of a sufficiently strong shock wave that creates four sectors with cracks, where the equivalent (shear) stress is reduced. Outside these sectors, the material behaves elastically, as shown by the cruciform area of high residual shear stress in the middle Fig. 5A, where the color centers previously created by the XFEL beam can be erased. Thus, the concentration of color centers should remain high in four spots according to the experimental distribution in the right Fig. 5A. The observed increased brightness of the thin rim around the cavity is probably due either to rapid crystallization of the melt on the cold walls of the channel or to a very high concentration of damage leading to amorphization of the crystal.

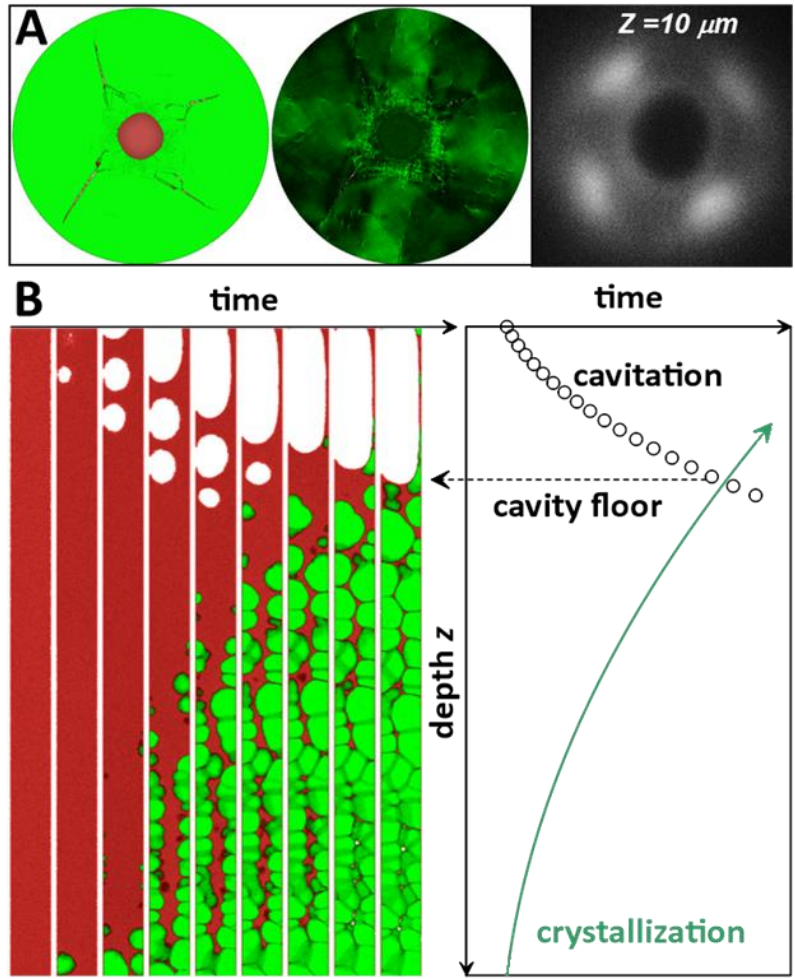

Fig. 5 **MD simulation results.** **A** — Simulations for $\xi_l$ = 297 kJ/cm$^3$ for the Zone 2 region (see Fig. 3A): Left cylinder – central symmetry parameter (green – crystal, red – disordered atom structure); middle – von Mises stress (dark green – lower von Mises stress after relaxation by damage); right – experimental image of LiF at a depth of $z$ = 10 μm (immediately after plugging). **B** — Simulations for $\xi_l[z = 1100$ μm$]$ = 12 kJ/cm$^3$, corresponding to the level at which the cavities $z \sim 1100$ μm are disappeared (see zone 3 in Fig. 3A). Full-scale simulation is presented in the corresponding Supplemental Video.

We have also performed MD simulations of processes leading to formation of cavity, which are driven by the slow loss of heat from the hot melt in the cylindrical channel into the surrounding cold crystal. At this stage, the pressure in the channel decreases with the melt due to cooling, and the radius of the channel remains almost unchanged as elastic stresses develop in the surrounding solid material, preventing the channel from narrowing. This means that the pressure in the melt can become negative even before it freezes, which creates the conditions for cavitation with the formation of cavities in the channel.

Figure 5B (left) shows the processes of crystallization and cavitation in the cylindrical channel with the fixed solid walls, through which the heat flows simultaneously over the entire length of the channel. A video of this MD simulation can be found in the Supplementary Materials. In the deep part, the melt has a lower temperature than in the upper part, so that crystallization starts at the bottom and spreads upwards in the channel. The speed of movement of the crystallization front is controlled by the degree of undercooling of the melt and can, in principle, exceed the speed of sound if the heat is lost to the channel walls fast enough. However, the cooling rate is slow, so the crystallization process proceeds below the speed of sound. Cavitation begins when the negative pressure in the melt falls below the tensile strength of liquid. This strength decreases with increasing temperature, so that cavitation propagates from top to bottom along the channel towards the crystallization front - as shown in the schematic $z$-$t$ diagram in Fig. 5B. We believe that the sharp cavity boundary deep in the LiF crystal seen in Fig. 3B is caused by the cessation of cavitation upon hitting the crystallization front.

*Estimation of the threshold for crater and cavity formation in a LiF sample*

To determine the threshold for crater formation and cavity opening, two-dimensional simulations were carried out using a hydrodynamic code for absorbed energy densities $\xi$ from 10.5 kJ/cm$^3$ to 28 kJ/cm$^3$ with a step of 3.5 kJ/cm$^3$ (model details in Methods).

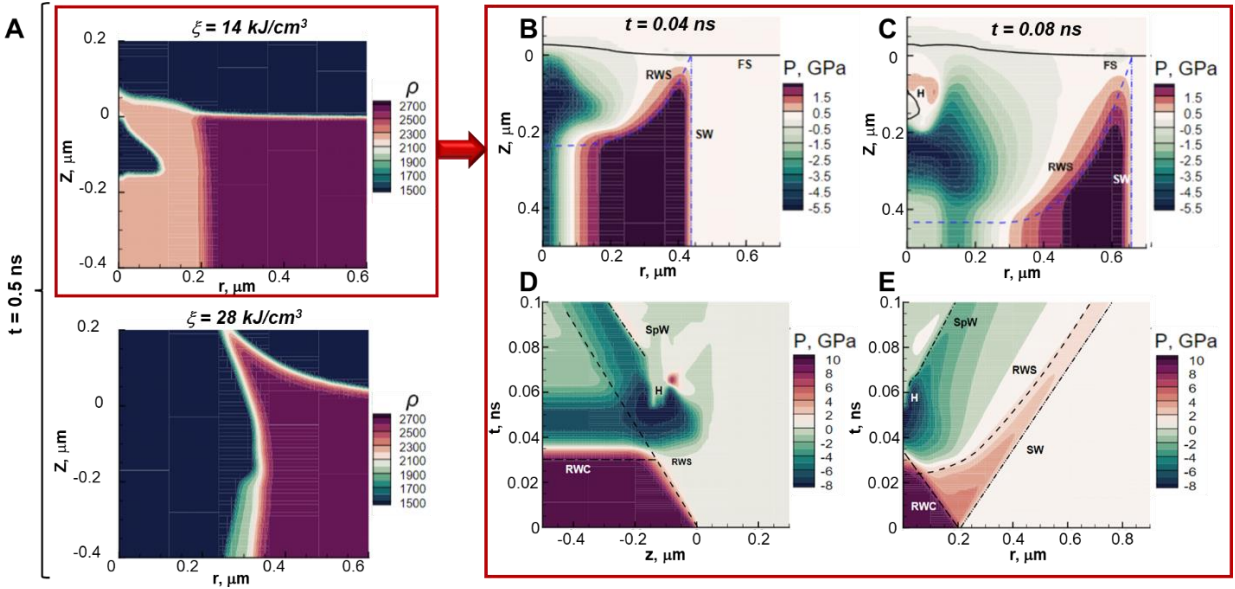

Fig. **6 Results of 2D HD modeling to investigate the threshold for manifestation of ablation and cavity. (A)** ─ Change in crater structure as a function of irradiation type at absorbed energy density of 14 kJ/cm$^3$ (upper frame) and $\xi$ = 28 kJ/cm$^3$ (lower frame) at a time of 0.5 ns after femtosecond XFEL irradiation; **(B-D)** Two-dimensional pressure maps in the crater region at the formation threshold $\xi_{crat}$ = 14 kJ/cm$^3$ for times **(B)** ─ 0.04 ns and **(C)** ─ 0.08 ns (**FS** – free surface, **SW** – radial shock wave, **RWS** – rarefaction wave emanating from the surface, **H** is the resulting cavity). **(D,E)** Overlap of surface **RWS** and radial **RWC** rarefaction waves. Space-time diagrams of the cratering process at $\xi_{crat}$ = 14 kJ/cm$^3$ – **(D)** z-t diagram on the plane $r = 0$ (symmetry axis of the XFEL beam) and **(E)** ─ r-t diagram on the plane $z = -0.15$ (depth of initial cavity formation). **RWS** – rarefaction wave from the surface of the material, **RWC** – radial rarefaction wave, **SW** – radial shock wave, **H** – cavity in the material, **SpW** – spall pulse).

The first signs of the formation of a surface structure leading to the formation of a crater were observed at an irradiance of $\xi_{crat}$ = 14 kJ/cm$^3$, Fig. 6A (top image). The calculated threshold for crater formation in LiF is generally consistent with the results of work[25]. Figure 6A also clearly shows that, in addition to the change in crater shape, a continuous cavity is formed at an irradiance $\xi_{cavity}$ of 28 kJ/cm$^3$, which is slightly higher than the experimental value found in Figure 3C - 12-18 kJ/cm$^3$. This could be since the simulations were performed for a layer close to the LiF surface (free boundary ─ LiF vacuum), while the clogging of the cavity is observed at a depth of $z \sim$ 1000-1100 μm.

Let us take a closer look at the process of crater formation. Figures 6(B-D) show the dynamics of crater formation when irradiated with a threshold value $\xi_{crat}$ = 14 kJ/cm³. Fig. 6(B,C) shows the dynamics of pressure development in the crater area at times of 0.04-0.08 ns. A discontinuity occurs in the thickness of the material, approximately at the height of $z$ = 0.15 μm from the initial free surface (**FS**). The line **FS** denotes the deformed boundary of the material, **RWS** is a hemispherical rarefaction wave propagating from the surface into the substance, the line **H-** denotes the boundary of the cavity formed in the material. We argue that the detachment of the material leading to the formation of a crater is caused by the addition of rarefaction waves coming from the surface of the material and from the boundary of the heated zone. The resulting tensile stresses lead to the formation of a crater. To illustrate this statement, consider two space-time diagrams: an *r-t* diagram on the plane $z = -0.15$ and a *z-t* diagram on the plane $r = 0$, Fig. 6(D,E). In Figures 6D and 6E, the usual dotted line shows the **RWS** rarefaction wave propagating from the free surface deep into the substance. At the beginning of its motion it is a plane wave, orthogonal to the axis of symmetry, but with time it acquires a curvature. It can be observed that the **RWS** and **SW** waves touch at the **FS** boundary of the sample (see Fig. 6B,C), which determines the curvature of the rarefaction wave.

Already at the time 0.4 ns behind the front of this wave there is an area with considerable negative pressure, about -6 GPa. This fact is related to the arrival of the second rarefaction wave, marked with **RWC** and a long-dotted line, at the same point in space. This rarefaction wave is associated with the decay of the discontinuity at the boundary of the heating jet in the radial direction along the entire heating depth. An element of the same decay is a shock wave **SW** propagating radially from the axis of symmetry, which has no influence on the processes taking place in the region of interest to us. The region of tensile stresses created by the addition of the waves eventually leads to the loss of continuity of the substance, i.e. spallation, denoted by **H**, occurs. After some time, during which the transient processes associated with the loss of continuity take place, spallation pulses occur, which are labeled with the dashed signature **SpW**. The spallation pulse propagating towards the free surface is not shown due to the complexity of its profile. After the spallation pulse has passed, the pressure in the resulting cavity gradually drops to almost zero.

## Discussion

The effect of the formation of a deep cylindrical submicron cavity observed in our work is new. We emphasize that the cavity is formed by the motion and deformation of a solid body in the depths of the condensed target ─ i.e. the mass of the vapor/plasma ejected from the target into the vacuum is much smaller than the mass of the substance that filled the cavity before laser irradiation. The results of this work can therefore be used to develop a method to create highly asymmetric, extended structures with submicron diameters in a LiF crystal using femtosecond X-ray pulses. The method consists of "*extrusion*" a cavity in the sample with an XFEL beam using the laser method by placing a focal point on the surface of the crystal. The surface morphology of the resulting hole can be observed with a scanning electron microscope. Monitoring the internal structure of the cavity with a non-destructive method ─ using a confocal laser scanning microscope by recording the presence or absence of a fluorescent signal within the beam image.

It is worth noting that structures in the form of extended cavities should form when irradiated with X-ray photons with energies of several keV not only in LiF, but also in ceramics, semiconductors and metals. Aluminum, for example, has similar mechanical properties to LiF such as the cold curve (i.e. pressure versus density at zero temperature), the shock Hugoniot as well as the bulk modulus and density. For the Si semiconductor, a similar structure in the form of an extended cavity (aspect ratio 1:10) was observed in the work[31] when it was irradiated with focused XFEL pulses ($d_{FWHM}$ = 1 μm, $\tau$ = 20 fs) with a photon energy of 10 keV and an absorbed energy density $\xi$ = 430 kJ/cm$^3$/pulse. The damage thresholds of Si and LiF are similar (5.82 kJ/cm$^3$/pulse and 4 kJ/cm$^3$/pulse), and the densities of these materials are close (2.65 and 2.64 g/cm$^3$). The authors of the paper[31] assumed that this structure was formed by the ejection of matter from the channel along the propagation axis of the XFEL beam. However, our work is the first to reveal the mechanism of formation of such long cylindrical cavities in solid materials ("radial extrusion") under the influence of hard X-ray photons. Thus, methods for forming deep, thin cavities will certainly be used in laser technologies for processing solids.

Another important area is the investigation of polymorphic (allotropic) transformations in the solid phase. Ultrashort optical laser-induced microexplosion in confined geometry has already demonstrated the formation of previously unknown high-pressure material phases such as two new energetically competitive tetragonal structures of silicon[32,33]. Other examples include switching the valence in Fe-atoms in olivine[34] the spatial separation of ions with different mass and formation of molecular oxygen inside the voids in oxides[35] and the formation of N-vacancies in c-BN crystals[36]. All these transformations occurred in near-spherical shaped nanovolumes contained within a thin surface layer. Estimates from transmission electron microscopy (TEM) images [23] suggest that the amount of new phase produced in a single optical laser shot is of the order of ~$10^{-15}$ g – $10^{-14}$ g, which makes it extremely difficult to further characterise its electronic properties Ultrashort XFEL pulses in hard, several keV, X-ray energy open up completely unique possibilities for highly targeted laser restructuring of materials. The presented results constitute a robust benchmark for the formation of broad range of exotic high-pressure material with extraordinary properties and to studies of highly non-equilibrium electron and ion dynamics in warm dense matter.

Note that for the development of the applications described above, further experiments are required to investigate the dependence of the cavity depth and its diameter on the energy of the incident photons, the impact energy, the size of the incident beam and the sample material.

## Materials and Methods

### Target details and conditions of exposure

Circle LiF crystal of 2 mm thickness and of 2000 mm diameter (two-side polished with $R_z$ = 0.05 um) was used as target sample. To ensure consistency and reproducibility, each exposition was performed on a fresh sample location.

We investigated two single-pulse irradiation regimes of a LiF crystal with pulse energies $E_1$ = 26.9 µJ/pulse and $E_2$ = 81 µJ/pulse (10% error). In both regimes, the sample surface was at the point of best focus. The reduction in pulse energy was achieved by using a beam attenuator. To estimate the extent of the impact on the sample, we used the absorbed energy density per pulse $\xi$ on the LiF surface, calculated as:

$$\xi [z = 0] = \frac{E_{pulse}}{S * d_{att}} = \frac{E_{pulse}*4*\text{Ln}(2)}{\text{Pi}*d_{att}*d_{FWHM}^2}, \quad (1)$$

where $E_{pulse}$- pulse energy, $d_{att}$- attenuation length in matter, $d_{FWHM}$– beam size at half maximum.

Considering the beam size and the attenuation length $d_{att}$ = 475 µm of 9 keV photons in LiF [37], the absorbed energy density was $\xi_1$ = 297 kJ/cm³ and $\xi_2$ = 895 kJ/cm³ on the sample surface. The attenuation of the beam as it propagates in LiF is due to two factors. The first is the absorption in the substance and the second is the attenuation due to the geometric divergence of the beam. Asymptotically, outside the waist zone, the beam is not a cylinder due to the broadening, but a cone with a very small but finite opening angle at the tip of the cone. The density of the absorbed energy ξ on the beam axis $z$ is the same:

$$\xi[z] = \xi_0[z=0]\exp(-z/d_{att})\frac{S_0}{S(z)}, \qquad (2)$$

here $\xi_0$ – density of absorbed energy on the target surface, i.e. at $z = 0$, $S_0$ and $S(z)$ – cross-sectional area of the beam at the surface and depth $z$ of the crystal, respectively.

The radius $r_{beam}(z)$ of the beam cross section at depth $z$ is equal to:

$$r_{beam}^2(z) = r_0^2\left[1 + \left(\frac{z}{z_R}\right)^2\right] = r_0^2\left[1 + \left(\frac{\ln(2)z\lambda M^2}{2\pi r_0^2}\right)^2\right], \qquad (3)$$

where $r_0$ is the spot radius of the beam at the focal plane, $z_R$ is the Rayleigh range, $\lambda$ is the wavelength of the photon and $M^2$ is the beam quality factor.

Using equations (2)-(3) and the parameter $M^2=3$, which we found earlier in work [25] for a given focusing of the XFEL beam, we plotted the dependence of the decay $\xi[z]$ in the LiF crystal, presented in Fig. 3C:

$$\xi[z] = \frac{\xi_0[z=0]}{1+\left(\frac{\ln(2)z\lambda M^2}{2\pi r_0^2}\right)^2}\exp(-z/d_{att}) \qquad (4)$$

**Investigation of cavity morphology**

After irradiation, the LiF crystal was examined using various readout systems, see Fig. 1B:

*Step I* - A scanning electron microscope (SEM) was used to examine surface damage – HELIOS NANOLAB 600I, FEI.

*Step II* - To obtain information about the intensity distribution of the incident beam inside the crystal, a Carl Zeiss LSM 700 confocal laser scanning microscope in fluorescence mode was used. Since the LiF crystal is a fluorescent medium, X-ray irradiation creates color centers (COs) in it, the distribution of which corresponds to the irradiation area[25,38,39]. The measurements were performed using an excitation laser with a wavelength of $\lambda = 488$ nm and an objective with a magnification of 100$^x$. The signal was recorded layer by layer with a layer thickness of $\Delta z = 1$ µm at a spatial resolution of 0.25 µm (field of view 200 × 200 µm$^2$).

*Step III* - For several test points, the FIB-SEM (Focused Ion Beam Scanning Electron Microscopy, FEI Helios NanoLab 600I) technology was used, which made it possible to examine the structure of the destruction area in its cross-section, i.e. its depth. In contrast to an electron microscope, the FIB inherently destroys the sample. When the high-energy ions hit the sample, they blast atoms off the surface. The FIB tool was used to etch the LiF sample layer by layer in the area next to the damage, while a scanning electron microscope was used to visualize the morphology of the damage (perpendicular to the XFEL beam propagation plane). The preparation of the sample before milling consisted in a thin gold coating (few nm) on the whole surface, followed by a thin protective layer of platinum locally deposited on the top of the hole. The platinum layer can be seen in the cross-section SEM images at the hole entrance. Then FIB milling was performed varying the current from nA (for opening a viewing section) to pA (for precise slicing of the hole).

**Simulations**

*2D/3D SPH*

Our 2D SPH code equipped with the LiF damage model [25] was used to solve the problem of cavity expansion in the Zone1-2 regions (see Fig. 3A). Since the heating zone of the incident XFEL beam in the material acts on the cold solid material like a gas piston at pressure changing with time, the problem can be simplified by replacing the hot material zone with a boundary condition for the pressure $P(t)$ set at the boundary between hot and cold material. In SPH simulations, to account for the initial laser heating within a cylinder with a radius $r_{beam} = 205$ nm, a uniform heating is given by the specific energy $e(z) = \xi(z)/\rho$, calculated according to eq. (4) at a certain depth $z$ from the sample surface.

The instantaneous heating due to the absorption of a laser pulse is accompanied by a pressure jump of $\Delta P(z) = (\gamma - 1)\rho_0 e(z)$, where $\gamma \simeq 1.75$ is the adiabatic exponent, $\rho_0 = 2.65 \text{g/cm}^3$ is the nominal density of LiF. Such localized heating leads to the formation of a strong shock wave, which propagates through the "cold" material (outside the heating area) and leads to its damage. However, it is worth noting that the amplitude of the shock wave decreases quite rapidly with moving away from the heating area, which limits the damage area to about ten microns or less, see insets (F2, F3) to Fig. 4.

To analyze the dynamics of the cavity opening near the surface (zone 1 in Fig. 3A), the modeling was performed in a 3-dimensional environment, for which. The SPH code does not use the axial symmetry of the flow. Thus, although the near-surface flow has two spatial coordinates (r, z), the simulation uses SPH particles that are initially arranged in a grid in the Cartesian (x, y, z) coordinate system – accordingly, this is referred to as 3D modeling. In contrast to the expansion of the cavity in depth, here deformations are mainly determined by the unloading of the resulting stresses nearby the free surface.

*MD*

MD simulations of LiF were performed with a newly developed pairwise potential for interaction between Li+ and F- ions in the following form:

$$V_{ij}(r) = \left[\frac{q_i q_j \exp(-r/d)}{r} + \frac{a_4}{r^4} + \frac{a_6}{r^6} + \frac{a_8}{r^8}\right] f(r, r_c), \quad (5)$$

where $q_i$ and $q_j$ are the known electrical charges of o*i*- and *j*-ions, while $d, a_4, a_6, a_8$ are fitting parameters.

Smoothing function $f(r, r_c)$ causes the above potential goes to zero with an interatomic distance $r$ approaching a cutoff radius $r_c$. The shielding with length $d$ is used to escape time-consuming calculations of the long-range Coulomb forces. The potential parameters were fitted to the known cold pressure curve [40,41] in a wide range of compression and stretching by using the stress-matching method [42]. The obtained potential provides the melting temperature of 990 K and reproduces the experimental difference of 25% between the densities of molten and solid LiF at the melting point. All MD simulations were performed with our in-house parallel MD code MD-VD$^3$ [43]. Since even the large-scale MD sample size is limited by a micrometer, our MD simulations are intended to obtain qualitative description of basic processes leading to formation of cavity in depth of LiF crystal heated to relatively low temperatures.

Simulation results presented in Fig. 5A were obtained after fast heating during 200 fs of a cylindrical sample with radius of 300 nm (in *XY* plane) and thickness of 8 nm along *Z*-axis with periodical conditions. The heating was performed by the Langevin thermostat using a target Gaussian-like temperature profile in a central spot with a radius of 30 nm. The peak temperature of 19 kK and pressure of 54 GPa are generated at the central axis.

Simulation results presented in Fig. 5B were obtained during long-term cooling of molten LiF in a cylinder with length of 1000 nm surrounded by the rigid walls at the radius of 10.3 nm. The initial sample was obtained by melting of solid LiF with a linear temperature gradient from 1300 K at the bottom of cylinder and 1500 K at the top. The pressure of 1.8 GPa was established along the cylinder, which is close to that remained after stop of material movement in MD simulation of early stage of evolution shown in Fig. 5A. Then the Langevin thermostat was applied on x- and y-velocity of atoms to cool the initial temperature distribution by 900 K to a target slope of 300-500 K along *Z*-axis with the characteristic cooling time of 200 ps. Figure 5B shows only part of sample nearby the cavity floor. See the full-scale simulation in the corresponding Supplemental Video.

*2D HD*

To estimate the thresholds for cratering and cavity opening, a computational algorithm based on Baer-Nunziato's system of equations [44], which uses a hydrodynamic approach to describe multiphase media, was used. Of course, the use of approximate equations for the state of aggregation and the lack of consideration of elastoplastic effects significantly limit the scope of the model for this type of problem, but the results obtained can be correlated qualitatively and quantitatively with experimental observations and computational results from other methods. The problem of the interaction of a rectangular beam with a LiF layer, without damping along the radius and deep into the substance, was considered.

To determine the threshold for crater formation and cavity opening, simulations were performed with irradiances in the range of absorbed energy densities of 10.5 kJ/cm$^3$ - 28 kJ/cm$^3$ with a step of 3.5 kJ/cm$^3$. A uniform rectangular grid consisting of 500 cells in the r-direction and 300 cells in the z-direction was used for the simulations. The symmetry condition is used on the left edge of the computational domain and non-reflecting boundary conditions are used for the rest. The dimensions of the computational domain were chosen so that possible perturbations of the wave pattern by non-reflecting boundary conditions do not affect the region of interest of the medium: 5 μm along the radius and 30 μm along the axis. For the numerical solution of the mathematical model, a variant of the HLLC method for the Baer-Nunziato equations was used, which is described for example in[45], where it was used to simulate high-speed plate collisions. This method has proven successful in solving problems with explicit contact boundaries, in contrast to the well-known HLL method for the Baer-Nunziato equations, which is characterized by a significant circulation viscosity at the contact boundary.

**Author contributions:**

Conceptualization: SSM, SAP, UZ

Methodology: SSM, SAP, TAP

Conducting an experiment: SSM, MM, MN, TRP, KA, ZK, VC, EB, JPS, IM, VV, VH, TB, JC, LJ, UZ, SP

Investigation: SSM, SAG, VVZ, TAP, NAI, VAK

Visualization: SSM, SAG, VVZ, PC, NO

Readout procedure: SSM, TAP, TS, AT, LR, AVR, SJ

Modeling: SAG, NAI, VAK, VVZ, YuVP, VS, PC, EP

Writing—original draft: SSM

Writing—review & editing: TAP, NAI, VVZ, SAP, LJ, VH, AVR, LR, RK

**Competing interests:**

Authors declare that they have no competing interests.

**Data and materials availability:**

All data are available in the main text or the supplementary materials. The data that support the findings of this study are available from the corresponding authors upon request. The HD, SPH and MD codes used for this study are available on reasonable request to petchu@mail.ru, grigorev@phystech.edu and 6asi1z@gmail.com respectively.